\documentclass[aps,pra,twocolumn]{revtex4}
\usepackage[dvips]{epsfig}
\usepackage[usenames]{color}
\usepackage{pstricks}
\usepackage{amsmath}
\usepackage{amssymb}
\usepackage{subfigure}
\usepackage{afterpage}

\begin{document}

\title{Tuning ultracold collisions of excited rotational dipolar molecules}

\author{Gaoren Wang, Goulven Qu{\'e}m{\'e}ner}
\affiliation{Laboratoire Aim{\'e} Cotton, CNRS, Universit{\'e} Paris-Sud, ENS Cachan, Campus d’Orsay, B{\^a}t. 505, 91405 Orsay, France}

\date{\today}

\begin{abstract}
We investigate the ultracold collisions of rotationally excited dipolar molecules in free-space, taking the hetero-nuclear bi-alkali molecule of KRb as an example. We show that we can sharply tune the elastic, inelastic and reactive rate coefficients of lossy molecular collisions when a second rotationally excited colliding channel crosses the threshold of the initial colliding channel, with the help of an applied electric field, as found by Avdeenkov et al. for non-lossy molecules [Phys. Rev. A 73, 022707 (2006)]. We can increase or decrease the loss processes whether the second channel is above or below the initial channel. This is seen for both bosonic and fermionic molecules. Additionally, we include the electric quadrupole and octopole moment to the dipole moment in the expression of the long-range multipole-multipole interaction. We found that for processes mediated by the incident channel like elastic and loss collisions, the inclusion of quadrupole and octopole moments are not important at ultralow energies. They are important for processes mediated by state-to-state transitions like inelastic collisions.
\end{abstract}


\maketitle

\font\smallfont=cmr7

\section{Introduction}

The recent progresses in the quantum-controlled preparation of dipolar molecules~\cite{Quemener_CR_112_4949_2012} have led to a strong interest in the ultracold scientific community. As quantum-controlled polar molecules can be used for many important applications~\cite{Carr_NJP_11_055049_2009}, it is crucial that experimentalists produce dense and long-lived species of them.
Two main mechanisms can prevent long lifetimes of ultracold molecules. The first one is chemical reactivity~\cite{Zuchowski_PRA_81_060703_2010,Byrd_PRA_82_010502_2010,Meyer_PRA_82_042707_2010}. The two initial reactants are transformed to two chemical products with a release of kinetic energy bigger than the depth of the trap, resulting in loss of molecules.
If no chemical reactivity is present, the second mechanism (yet not experimentally observed) comes from the possibility for non-reactive molecules to long enough explore the large phase-space density of the molecule-molecule complex. The transient complex lives long enough to collide with a third molecule resulting again in loss of molecules~\cite{Mayle_PRA_87_012709_2013}.
To overcome this, one can use the anisotropy of the dipole-dipole interaction of dipolar molecules but this requires to add a one-dimensional (1D) confining optical lattice to protect the molecules from coming close~\cite{Quemener_PRA_81_060701_2010,Micheli_PRL_105_073202_2010,Quemener_PRA_83_012705_2011,DeMiranda_NP_7_502_2011}
and usually strong confinement is needed.
But is there a way to protect those molecules even in a three-dimensional (3D) free-space environment? For example, OH molecules can have large elastic collisions compared to inelastic ones~\cite{Avdeenkov_PRA_66_052718_2002,Ticknor_PRA_71_022709_2005,Quemener_PRA_88_012706_2013}, and this comes from a repulsive van der Waals interaction~\cite{Stuhl_N_492_396_2012}.
Is there a similar mechanism that can be applied to ultracold heteronuclear alkali-based dipolar molecules which arouse a growing experimental interest?

In this study we propose to use excited rotational states and electric fields to tune the collisional properties of lossy dipolar molecules in free-space, taking fermionic~\cite{Ni_S_322_231_2008} and bosonic~\cite{Aikawa_PRL_105_203001_2010} KRb molecules as examples. We consider the full rotational structure in the quantum mechanical problem and
take into account the accurate long-range electronic van der Waals interaction.
We also consider the effect of the electric quadrupole and octopole moments besides the electric dipole moment in the expression of the long range multipole-multipole interaction.
This study is especially relevant for ultracold dipolar molecules produced in their first excited rotational state~\cite{Ospelkaus_PRL_104_030402_2010,Neyenhuis_PRL_109_230403_2012}
which are essential to a lot of applications such as tailoring interactions with external fields and confinements~\cite{Micheli_PRA_76_043604_2007}, suppressing of loss collisions with combination of static electric and microwave fields~\cite{Gorshkov_PRL_101_073201_2008}, quantum magnetism with polar molecules~\cite{Gorshkov_PRL_107_115301_2011,Gorshkov_PRA_84_033619_2011},
dipolar spin-exchange interactions
in optical lattices~\cite{Yan_N_501_521_2013,Hazzard_PRL_113_195302_2014} and suppression of molecular loss with continuous quantum Zeno effect~\cite{Zhu_PRL_112_070404_2014}.

The paper is organized as follow. Section 2 deals with the theoretical formalism we used in the paper. In Section 3 we present and discuss our scientific results. Finally we conclude in Section 4.

\section{Theoretical formalism}

We consider collisions of polar molecules in excited rotational state. We take the heteronuclear bi-alkali molecule of KRb~\cite{Ni_S_322_231_2008} as an example. We consider the molecules in their ground electronic X$^1\Sigma^+$ and vibrational $v=0$ state. Collisions occur in a three-dimensional free-space, under an applied electric field $\vec{E}$ which sets the axis of quantization $\hat{z}$. We use a time-independent quantum formalism using Jacobi coordinates to describe respectively the relative motion $\vec{\rho}_1$ of molecule 1, $\vec{\rho}_2$ of molecule 2, and the relative motion $\vec{r}$ between the center-of-mass of the two molecules. We consider only the rotational structure of the molecules, characterized by two usual quantum numbers $n$, and $m_n$ the corresponding projection of the rotational angular momentum onto the quantization axis. We note $|n, m_n \rangle \equiv Y_n^{m_n}(\hat{\rho})$ the corresponding eigenfunction of a rotating molecule. We do not consider the other internal structures as the electronic and the vibrational ones as the collision energy involved in this study is too low to excite molecules in higher electronic or vibrational state.
As the rotation - nuclear spin interaction ($\sim$Hz)
and the electric quadrupole interaction of the nuclei with the surrounding charges ($\sim$MHz)~\cite{Aldegunde_PRA_78_033434_2008}
are small compared to the rotational constant ($\sim$GHz), we neglect the nuclear spin structure in this study.

\begin{figure} [h]
\begin{center}
\includegraphics*[width=6.5cm,keepaspectratio=true,angle=-90]{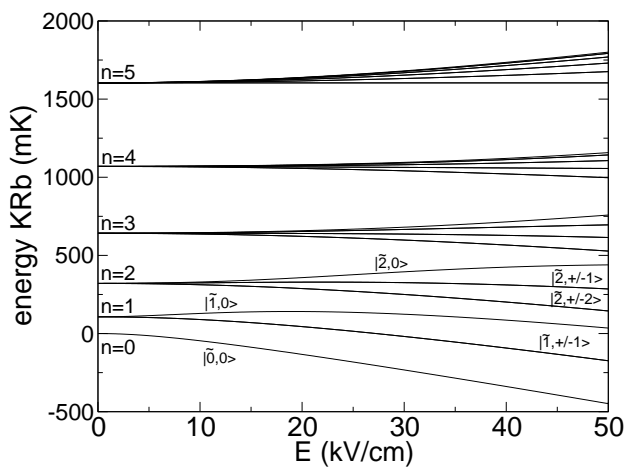}
\end{center}
\caption{Energy of $^{40}$K$^{87}$Rb as a function of the electric field. At zero electric field, the rotational quantum number $n$ is a good quantum number.}
\label{NRG-1PLE-FIG}
\end{figure}

The energy of a single molecule without electric field is provided by
the rotational term $ B_{\rm{rot}} \, n(n+1)$. We use the experimental value of the rotational constant $B_{\rm{rot}} = 1.113950$~GHz from Ospelkaus et al.~\cite{Ospelkaus_PRL_104_030402_2010} for the fermionic $^{40}$K$^{87}$Rb molecule and
$B_{\rm{rot}} = 1.095362$~GHz from Aikawa et al.~\cite{Aikawa_PRL_105_203001_2010} for the bosonic $^{41}$K$^{87}$Rb molecule.
In an electric field, we add the Stark term $ V_{S} = - \vec{d} \cdot \vec{E} $
where $\vec{d}$ is the full electric dipole moment in the body-fixed frame and $\vec{E} = E \, \hat{z}$ is the electric field.
In the basis set $|n, m_n \rangle $, the Stark term is written
\begin{multline}
 \langle n \, m_n | V_{S} | n' \, m_n' \rangle    = 
 - d \, E  \ \delta_{m_n,m_n'} \, (-1)^{m_n}  \\  \sqrt{2n+1} \, \sqrt{2n'+1} 
\, \left( \begin{array}{ccc} n & 1 & n' \\ 0 & 0 & 0 \end{array} \right)
\, \left( \begin{array}{ccc} n & 1 & n' \\ -m_{n} & 0 & m_{n}' \end{array}  \right) .
\label{VS}
\end{multline}
We use the experimental value of $d=0.574$~D for the full dipole moment of the KRb molecule~\cite{Ni_PhDThesis_2009}.
We diagonalize this matrix using a basis set including $n=[0-5]$ and get the dressed states $|\tilde{n}, m_n \rangle $ for a given electric field. The quantum number $m_n$ is conserved. The tilde corresponds to a certain admixture of different rotational quantum numbers due to the electric field but when $E \to 0$, the dressed states $|\tilde{n}, m_n \rangle$ correlate to the bare states $|n, m_n \rangle $. The corresponding energies of the dressed states $E_{\rm{dress}}$  are shown in Fig.~\ref{NRG-1PLE-FIG} for the fermionic $^{40}$K$^{87}$Rb molecule, for $E=[0-50]$~kV/cm. Note that with the actual experimental set-ups the value of $E \sim 30$~kV/cm corresponds already to an upper experimental value of the electric field. We got similar results for the bosonic $^{41}$K$^{87}$Rb molecule.

\begin{figure} [h]
\begin{center}
\includegraphics*[width=6.5cm,keepaspectratio=true,angle=-90]{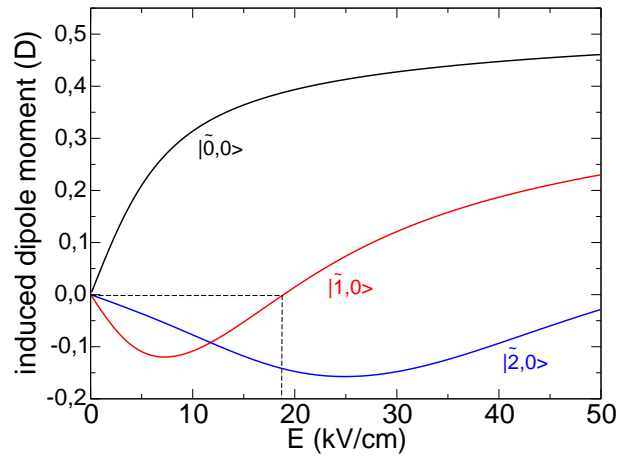}
\end{center}
\caption{(Color online) Induced dipole moment of KRb as a function of the electric field for the three states $|\tilde{0},0\rangle$ (black curve), $|\tilde{1},0\rangle$ (red curve), $|\tilde{2},0\rangle$ (blue curve). The induced dipole moment for the state $|\tilde{1},0\rangle$ vanishes at an electric field of $\sim 19$~kV/cm.}
\label{DVSE-FIG}
\end{figure}

It is also useful to plot the induced dipole moment along the electric field direction in the space-fixed frame. The induced dipole moment is the mean value of the dipole moment over the dressed state $|\tilde{n}, m_n \rangle$ at a given electric field $E$
\begin{eqnarray}
d_{\rm{ind}}(E_0) = \langle \tilde{n}, m_n \, | \, \vec{d} \cdot \hat{z} \, | \, \tilde{n}, m_n \rangle \, \bigg|_{E_0} = - \frac{dE_{\rm{dress}}}{dE} \, \bigg|_{E_0} .
\end{eqnarray}
This is shown in Fig.~\ref{DVSE-FIG} for the fermionic $^{40}$K$^{87}$Rb molecule. The ground rotational state $|\tilde{0}, 0 \rangle $ has a positive induced dipole moment growing in a monotonic way from 0 to $d=0.574$~D. The induced dipole moment of the second lowest rotational state $|\tilde{1}, 0 \rangle $ is first negative and then becomes positive at $E \ge 19$~kV/cm, where the derivative of the dressed energy is zero in Fig.~\ref{NRG-1PLE-FIG}. This is due to an interplay in the Stark effect of a negative contribution of $d_{\rm{ind}}$ from the $|0, 0 \rangle $ bare state which dominates in the range $E=[0-19]$~kV/cm and of a positive contribution of $d_{\rm{ind}}$ from the higher bare states $|2, 0 \rangle, |3, 0 \rangle, ...$. A similar behaviour is seen for the dressed state $|\tilde{2}, 0 \rangle $.
Also, we note that it is harder to induce an electric dipole moment with an electric field as the initial rotational quantum number increases. At a maximum experimental electric field of $E=30$~kV/cm, $d_{\rm{ind}} \sim 0.1$~D only for $|\tilde{n}=1, m_n=0 \rangle$ KRb molecules.

The potential energy between two molecules is given by an electronic van der Waals interaction
\begin{eqnarray}
V_{\rm{vdW}} = -C^{\rm{el}}_6/r^6
\end{eqnarray}
and a long-range interaction given by a multipole-multipole expansion~\cite{Stone_Book_1996}
\begin{multline}
 V_{\rm{mult}} = \frac{1}{4 \pi \varepsilon_0} \sum_{\lambda_1 \, \lambda_2 \, \lambda} \, \sum_{\omega_{\lambda_1} \, \omega_{\lambda_2}} \, \frac{Q_{\lambda_1  \omega_{\lambda_1}} \, Q_{\lambda_2  \omega_{\lambda_2}}}{ r^{\lambda+1}}  \, \delta_{\lambda,\lambda_1+\lambda_2} \\
\sum_{m_{\lambda_1} \, m_{\lambda_2} \, m_\lambda}  {\cal A}(\hat{\rho}_1,\hat{\rho}_2,\hat{r})
\end{multline}
with $\lambda  = \lambda_1 + \lambda_2$. The angular part is given by
\begin{multline}
{\cal A}(\hat{\rho}_1,\hat{\rho}_2,\hat{r})  = 
(-1)^{\lambda_1} \ \left( \frac{(2 \lambda_1 + 2 \lambda_2 + 1)!}{(2 \lambda_1)!(2 \lambda_2)!}  \right)^{1/2} \\
\, \left( \begin{array}{ccc}  \lambda_1 & \lambda_2 & \lambda \\ m_{\lambda_1} & m_{\lambda_2} & -m_\lambda \end{array}  \right)  \\
 [D^{\lambda_1}_{m_{\lambda_1} \omega_{\lambda_1}} (\hat{\rho}_1)]^* \ [D^{\lambda_2}_{m_{\lambda_2} \omega_{\lambda_2}}(\hat{\rho}_2)]^*  \  [D^{\lambda}_{-m_{\lambda} 0}(\hat{r})]^*
\end{multline}
with $m_\lambda = m_{\lambda_1} + m_{\lambda_2} $.
$Q_{\lambda_i \omega_{\lambda_i}} $ is a generalized multipole in the body-fixed frame of the molecule. $\lambda_i$ is an angular momentum quantum number corresponding to the electronic charge distribution in the molecules $i=1,2$. $\lambda_i=0,1,2,3$ correspond respectively to the charge, the dipole, the quadrupole and the octopole moment. $\omega_{\lambda_i}=[-\lambda_i,+\lambda_i]$ is the projection of this angular momentum onto the body-fixed frame molecular axis. For $\Sigma$ electronic molecules, $\omega_{\lambda_1}=\omega_{\lambda_2}=0$.
We take $C^{\rm{el}}_6 = 12636$~E$_h$.a$_0^6$ corresponding to the value $C^{e}_6 + C^{g-e}_6$ of Lepers et al.~\cite{Lepers_PRA_88_032709_2013}.
We have already $ Q_{1  0}  \equiv d = 0.2258 \ \rm{a.u.} = 0.574$~D as mentioned above.
We used the value of Byrd et al.\cite{Byrd_PRA_86_032711_2012}, $ Q_{2 0} =15.14$~a.u. for the quadrupole moment and $ Q_{3  0} =-69.09$~a.u. for the octopole moment.

The total wavefunction is expanded into a symmetrized basis set of the internal dressed states of the molecules~\cite{Quemener_PRA_88_012706_2013}  and the external relative colliding motion
\begin{multline}
|\tilde{n}_1, m_{n_1}, \tilde{n}_2, m_{n_2}, l, m_l ; \eta \rangle  =   \frac{1}{\sqrt{2(1+\delta_{\tilde{n}_1,\tilde{n}_2}\delta_{m_{n_1},m_{n_2}})}}  \\
 \bigg\{ |\tilde{n}_1, m_{n_1} \rangle |\tilde{n}_2, m_{n_2} \rangle  + \eta \ |\tilde{n}_2, m_{n_2} \rangle |\tilde{n}_1, m_{n_1} \rangle  \bigg\} |l, m_l \rangle
\end{multline}
where $|l, m_l \rangle \equiv Y_l^{m_l}(\hat{r})$ is the spherical harmonics corresponding to the angular part of the relative colliding motion and $\eta=\pm1$ for respectively symmetric and anti-symmetric combinations of the initial rotational states. The symmetrization of the total wavefunction with interchange of indistinguishable particles implies that $\eta \, (-1)^l = \varepsilon$ where $\varepsilon=+1$ for indistinguishable bosons and $\varepsilon=-1$ for indistinguishable fermions. In this study, we consider indistinguishable particles with identical initial states so that only the symmetric combinations $\eta=+1$ are used. Then, for indistinguishable bosons only even values of $l=0,2,...$ are allowed and for indistinguishable fermions only odd values of $l=1,3,...$ are allowed. When we include only the dipole moment we use $l=0,2,4$ for the bosons and $l=1,3,5$ for the fermions. When we include the quadrupole and octopole as well, we use $l=[0-12]$ for the bosons and $l=[1-13]$ for the fermions.

The total wavefunction is then written
\begin{multline}
\psi(\vec{\rho}_1,\vec{\rho}_2, \vec{r}) =  \\  \frac{1}{r} \, \sum_{l} f_{l \, \tilde{n}_1 \, \tilde{n}_2}^{m_{\rm{tot}} \,\eta}(r)
\, |\tilde{n}_1, m_{n_1}, \tilde{n}_2, m_{n_2}, l, m_l ; \eta \rangle .
\label{Psi}
\end{multline}
Here $m_{\rm{tot}}=m_{n_1}+m_{n_2}+m_l$, and is conserved during the collision.
In this basis set, the van der Waals term is diagonal.
In the bare basis set $ |{n}_1, m_{n_1} \rangle |{n}_2, m_{n_2} \rangle |l, m_l \rangle$, the multipole-multipole interaction is given by
\begin{multline}
 \langle {n}_1, m_{n_1}, {n}_2, m_{n_2}, l, m_l  | V_{\rm{mult}} | {n}_1', m_{n_1}', {n}_2', m_{n_2}', l', m_l' \rangle  =  \\
 \frac{1}{4 \pi \varepsilon_0} \sum_{\lambda_1 \, \lambda_2 \, \lambda}
\, (-1)^{\lambda_1} \ \left( \frac{(2 \lambda_1 + 2 \lambda_2 + 1)!}{(2 \lambda_1)!(2 \lambda_2)!}  \right)^{1/2}
\, \frac{Q_{\lambda_1  0} \, Q_{\lambda_2  0}}{ r^{\lambda_1+\lambda_2+1}}   \\
  \sum_{m_{\lambda_1} \, m_{\lambda_2} \, m_\lambda}
 (-1)^{m_{n_1}+m_{n_2}+m_l}
 \, \left( \begin{array}{ccc}  \lambda_1 & \lambda_2 & \lambda \\ m_{\lambda_1} & m_{\lambda_2} & -m_\lambda \end{array}  \right) \\
\times  \sqrt{(2n_1+1) \, (2n_1'+1)}
\, \left( \begin{array}{ccc} n_1 & \lambda_1 & n_1' \\ 0 & 0 & 0 \end{array} \right)
\, \left( \begin{array}{ccc} n_1 & \lambda_1 & n_1' \\ -m_{n_1} & m_{\lambda_1} & m_{n_1}' \end{array}  \right)  \\
\times  \sqrt{(2n_2+1) \, (2n_2'+1)}
\, \left( \begin{array}{ccc} n_2 & \lambda_2 & n_2' \\ 0 & 0 & 0 \end{array} \right)
\, \left( \begin{array}{ccc} n_2 & \lambda_2 & n_2' \\ -m_{n_2} & m_{\lambda_2} & m_{n_2}' \end{array}  \right) \\
 \times  \sqrt{(2l+1) \, (2l'+1)}
\, \left( \begin{array}{ccc} l & \lambda & l' \\ 0 & 0 & 0 \end{array} \right)
\, \left( \begin{array}{ccc} l & \lambda & l' \\ -m_{l} & -m_{\lambda} & m_{l}' \end{array}  \right) .
\label{VMULT}
\end{multline}
As the symmetrized dressed states are expressed in terms of the bare states, we perform a transformation to get the expression of the multipole-multipole interaction in the symmetrized dressed states basis set.

\begin{figure} [h]
\begin{center}
\includegraphics*[width=6.5cm,keepaspectratio=true,angle=-90]{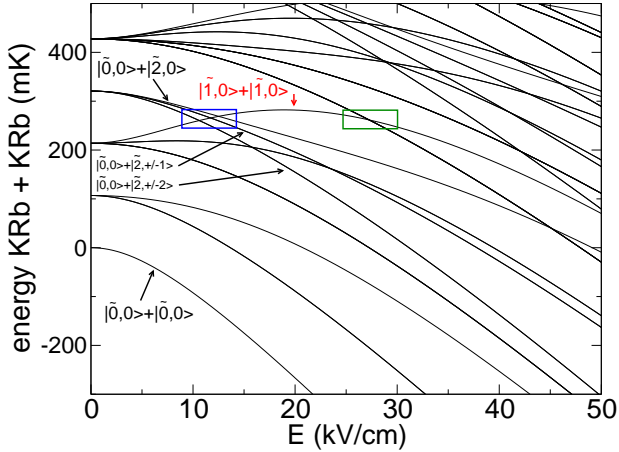}
\end{center}
\caption{(Color online)
Energy of $^{40}$K$^{87}$Rb + $^{40}$K$^{87}$Rb as a function of the electric field. The combined states in red correspond to the initial colliding states considered in this study. The blue box indicates an example of three crossings of different threshold levels, respectively $|\tilde{1},0\rangle$+$|\tilde{1},0\rangle$ with $|\tilde{0},0\rangle$+$|\tilde{2},\pm2\rangle$, $|\tilde{0},0\rangle$+$|\tilde{2},\pm1 \rangle$, and $|\tilde{0},0\rangle$+$|\tilde{2},0\rangle$ for increasing electric fields. The green box indicates the crossing of $|\tilde{1},\pm1\rangle+|\tilde{2},\pm2\rangle$ or $|\tilde{1},\pm1\rangle+|\tilde{2},\mp2\rangle$
with $|\tilde{1},0\rangle+|\tilde{1},0\rangle$.}
\label{NRG-2PLE-FIG}
\end{figure}

We solve the time-independent Schr\"{o}dinger equation
$ H \, \psi = E_{\rm{tot}} \, \psi $.
$H = T + V_{\rm{vdW}} + V_{\rm{mult}}$, where $T = - \hbar^2 \, \vec{\nabla}^2_{\vec{r}} / 2m_{\rm{red}}$ is the kinetic energy term and $m_{\rm{red}}$ the reduced mass of the system.
$E_{\rm{tot}} = E_{\rm{dress}_1} + E_{\rm{dress}_2} + E_c$ is the total energy and is the sum of the initial dressed state energy of molecule 1 and 2, and the collision energy $E_c$. The total energy is conserved during the collision.
The energy of the combined initial dressed states $E_{\rm{dress}_1} + E_{\rm{dress}_2}$ is shown in Fig.~\ref{NRG-2PLE-FIG} as a function of the electric field. In the rest of the study, we use a fixed collision energy of $E_c = 500$~nK and consider the collision of the combined initial dressed states $|\tilde{1},0\rangle$+$|\tilde{1},0\rangle$ indicated with a red arrow.

In this symmetrized basis set, we get a set of differential equations for the functions of the radial relative motion. We use a diabatic-by-sector method (see for example~\cite{Quemener_PRA_83_012705_2011}). We get the adiabatic energy curves by diagonalizing the term in Eq.~\ref{VMULT} and adding the diagonal centrifugal term $\hbar^2 \, l(l+1) / 2m_{\rm{red}}$ and the diagonal van der Waals term.
We solve the log-derivative of the radial functions for each value of the intermolecular separation $r$~\cite{Johnson_JCP_13_445_1973,Manolopoulos_JCP_85_6425_1986}. The log-derivative of the wavefunction is propagated from
$r_{\rm{min}}=10 \, a_0$ to $r_{\rm{max}}=10000 \, a_0$. At $r_{\rm{min}}$, we start with a log-derivative matrix that corresponds to a full absorbing potential at short-range~\cite{Idziaszek_PRA_82_020703_2010}. For reactive polar molecules~\cite{Zuchowski_PRA_81_060703_2010}, this accounts for a full loss at short-range due to the chemical reaction, while for non-reactive polar molecules, this accounts for possible long-lived sticky collisions~\cite{Mayle_PRA_87_012709_2013} that can be seen as loss processes in experiments. With this condition, only the background of the scattering is probed in this study with the electric field~\cite{Quemener_PRA_81_022702_2010,Idziaszek_PRA_82_020703_2010}, while the resonant part of the scattering is washed out and does not survive in the collision process.

Applying usual asymptotic boundary conditions at $r_{\rm{max}}=10000 \, a_0$ provides the observables such as the cross sections and rate coefficient as a function of the collision energy and the electric field for a given initial state of the molecules. \\

\section{Results. Collisions of $\rm{K}\rm{Rb}(|\tilde{1},0\rangle) +  \rm{K}\rm{Rb}(|\tilde{1},0\rangle)$}

\begin{figure} [h]
\begin{center}
\includegraphics*[width=6.5cm,keepaspectratio=true,angle=-90]{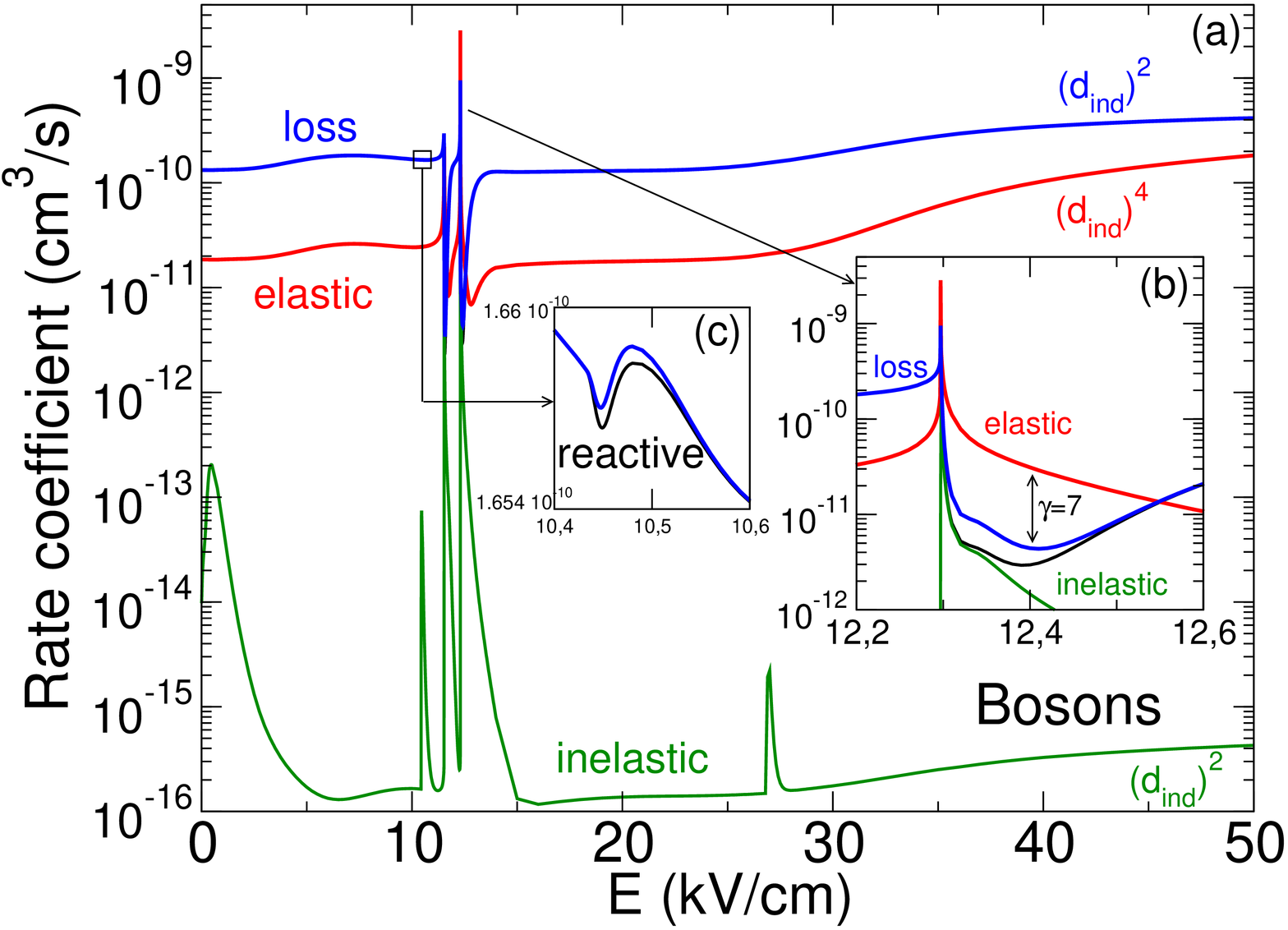} \\
\includegraphics*[width=6.5cm,keepaspectratio=true,angle=-90]{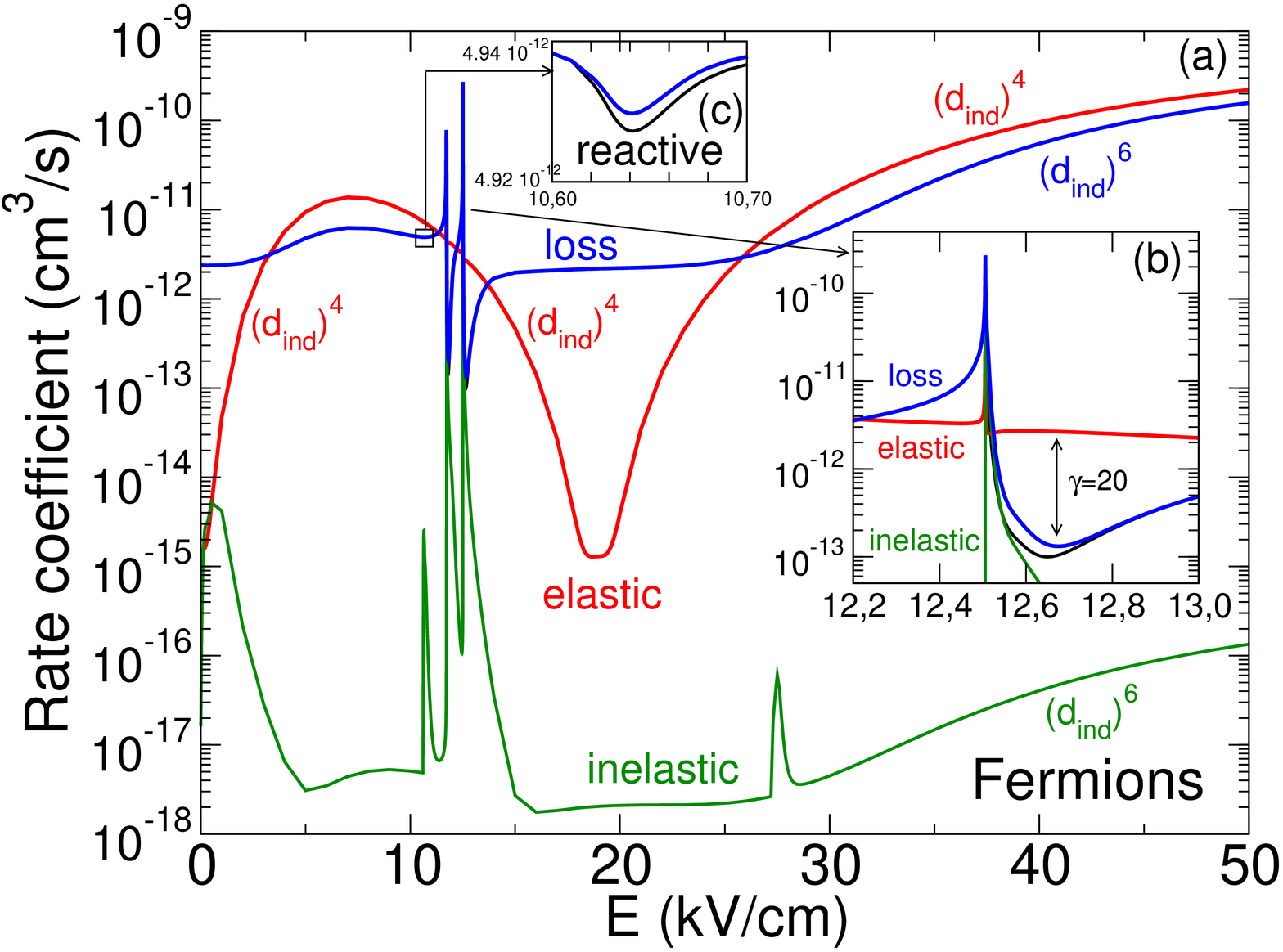}
\end{center}
\caption{(Color online) Rate coefficient for the initial colliding state $|\tilde{1},0\rangle$+$|\tilde{1},0\rangle$ at various electric fields. The collision energy is fixed to $E_c=500$~nK. Upper panel: bosonic $^{41}$K$^{87}$Rb. Lower panel: fermionic $^{40}$K$^{87}$Rb. Reactive: black curve. Elastic: red curve. Inelastic: green curve. Loss (= reactive + inelastic), blue curve. The reactive curves and loss curves are almost graphically indistinguishable.}
\label{RATEN1-FIG}
\end{figure}

We consider collisions of chemically reactive KRb molecules in the first rotational excited states $\tilde{n}=1,m_n=0$. We discuss here the behaviour of the rate coefficients for the elastic processes, reactive processes, and inelastic processes to other combined dressed states.
Avdeenkov et al.~\cite{Avdeenkov_PRA_73_022707_2006} have performed an earlier work on that subject for non-lossy molecules, using a non-absorbing repulsive hard wall at short-range and no electronic van der Waals interaction between the molecules. Knowing more now on the properties of ground state KRb molecules~\cite{Ospelkaus_S_327_853_2010,Ni_N_464_1324_2010}, we included an accurate electronic van der Waals coefficient~\cite{Lepers_PRA_88_032709_2013}, an absorbing potential to take into account their chemical reactivity~\cite{Zuchowski_PRA_81_060703_2010}
and tested the inclusion of electric quadrupole and octopole moments~\cite{Byrd_PRA_86_032711_2012}. We believe our results represent a more quantitative version of the pioneering work of Avdeenkov et al. and could be directly compared with experimental measurements.
We present in Fig.~\ref{RATEN1-FIG} the corresponding rate coefficients for bosons (upper panel) and fermions (lower panel) as a function of the electric field. The loss rates is the sum between the reactive and inelastic rates. The sum is almost graphically identical with the reactive one. 
Only the dipole has been included in the multipole-multipole interaction for the result presented on the graph. The effect of the quadrupole and octopole is presented in Appendixes B and C at selected values of the electric field. \\

Bosons and fermions show similar behaviours except that the bosonic rates are globally higher than the fermionic ones. This is expected from the even parity of the $l$ numbers for bosons which include the s-wave barrierless-mediated collisions in contrast with the odd $l$ numbers for fermions which include only barrier-mediated collisions.
Globally for bosons, the loss rate dominates over the elastic rate. For fermions, the loss and elastic rates have a much more similar magnitude except when the induced dipole moment is zero where reactive rates are bigger than elastic ones.
Looking at Fig.~\ref{DVSE-FIG}, the induced dipole moment of the $\tilde{n}=1,m_n=0$ state is zero at $E=0$~kV/cm and $E \sim 19$~kV/cm. For both fields, the reactive or the elastic rates share the same value, for the bosonic or the fermionic system.
In between, the absolute value of the induced dipole moment increases and decreases corresponding to the increase and decrease of the rates in this range. For $E > 19$~kV/cm, the induced dipole moment increases in a monotonic way corresponding to the monotonic increase of the reactive and elastic rates. 
By plotting the rates as a function of the induced dipole moment (see correspondence in Fig.~\ref{DVSE-FIG}), we found that the fermionic reactive and inelastic rates behave as 
$d_{\rm{ind}}^6$ as found in~\cite{Quemener_PRA_81_022702_2010} and that the bosonic reactive and inelastic rates behave as $d_{\rm{ind}}^2$ as found in~\cite{Quemener_PRA_84_062703_2011}
for $\tilde{n}=0,m_n=0$ collisions.
The fermionic and bosonic elastic rates behave as $d_{\rm{ind}}^4$ as found in~\cite{Bohn_NJP_11_055039_2009}. \\

In contrast with collisions between ground states $\tilde{n}=0,m_n=0$ molecules,
as recalled and discussed in Appendix A,
inelastic transitions to lower combined states (all the combined states below
$|\tilde{1},0\rangle+|\tilde{1},0\rangle$ in Fig.~\ref{NRG-2PLE-FIG}) can occur.
As the potential energy surfaces of the K$_2$Rb$_2$ complex are unknown in the present time, we do not know how the different combined states of the molecules couple at short range. In other words, inelastic collisions can be sensitive to what happens at short range. Knowing nothing about that, we chose an initial log-derivative matrix with no coupling between different combined molecular states. This initial choice will only underline the effect of inelastic transitions that occur at long range. Within this condition, the inelastic rate coefficients are globally small compared to the reactive and elastic ones.
In appendixes B and C, we show the convergence of the reactive, elastic and inelastic rates, when only the dipole is included in Eq.~\ref{VMULT} (case D), the dipole and quadrupole (case DQ) and finally the dipole, quadrupole, and octopole (case DQO).
This is performed for the bosons and fermions and for $\tilde{n}=0,m_n=0$ and $\tilde{n}=1,m_n=0$ molecular collisions.

Globally there is not significant effect on ultracold processes mediated by the incident channel like the reactive and elastic ones. 
On one hand, the characteristic length $a_s = (2 \, m_{\rm{red}} \, C_s /\hbar^2)^{1/{s-2}}$~\cite{Gao_PRA_78_012702_2008} of 
the dipole-dipole (D-D) interaction ($\lambda_1=\lambda_2=1, \lambda=2$), the quadrupole-quadrupole (Q-Q) interaction ($\lambda_1=\lambda_2=2, \lambda=4$) and 
the octopole-octopole (O-O) interaction ($\lambda_1=\lambda_2=3, \lambda=6$)  are respectively $a_3=11792$~a$_0$, $a_5=376$~a$_0$ and $a_7=64$~a$_0$, taking $C_s \sim Q_{\lambda_1  0} Q_{\lambda_2  0}$ as an upper limit of magnitude and $s = \lambda+1$.
On the other hand, the D-D interaction implies a coupling rule in Eq.~\ref{VMULT} of $|l-2| \le l' \le l+2$, the Q-Q interaction a coupling rule of $|l-4| \le l' \le l+4$, and the O-O interaction a coupling rule of $|l-6| \le l' \le l+6$.
So, higher multipole-multipole interactions become important at shorter distances as their characteristic length decreases. However higher multipoles imply couplings with higher values of $l$ or $l'$, meaning higher centrifugal barriers. For ultralow collision energies in the incident channels, those barriers prevent the particles to reach those short characteristic distances. Therefore the quadrupoles and octopoles do not play a major role for ultracold elastic and reactive processes compared to dipoles.
This condition is removed when the collision energy is large enough
to overcome those centrifugal barriers and the molecules can freely access the short-range region where the high multipolar interactions are characteristic. This is the case if the incident collision energies are increased (not studied here), leaving the ultracold regime. This is also the case for collision processes of ultracold molecules mediated by a change of states like the inelastic ones (studied here) for which the final kinetic energies are high enough to explore the short-range physics. Therefore the quadrupoles and octopoles have a stronger effect on ultracold inelastic state-to-state transitions as seen in the tables of Appendixes B and C. \\

\begin{figure} [h]
\begin{center}
\includegraphics*[width=9cm,keepaspectratio=true,angle=0]{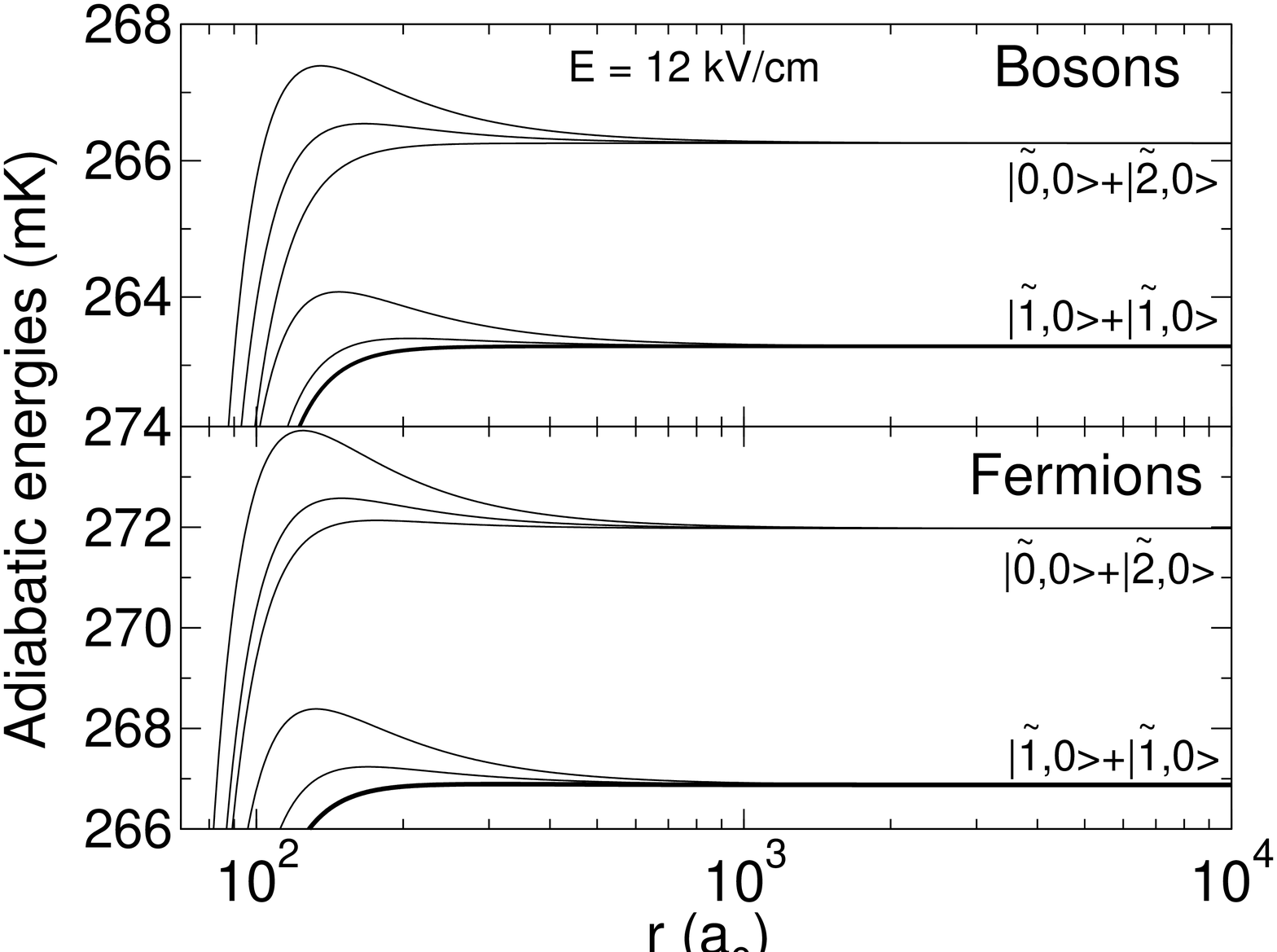}
\end{center}
\caption{Adiabatic energies for the initial colliding state $|\tilde{1},0\rangle$+$|\tilde{1},0\rangle$ at a fixed electric field of $E=12$~kV/cm. Upper panel: bosonic $^{41}$K$^{87}$Rb. Lower pannel: fermionic $^{40}$K$^{87}$Rb. The intital colliding state is just below the coupled state $|\tilde{0},0\rangle$+$|\tilde{2},0\rangle$ for this electric field. The black bold curve represents the lowest incident colliding channel.}
\label{SPAG-FIG}
\end{figure}

\begin{figure} [h]
\begin{center}
\includegraphics*[width=9cm,keepaspectratio=true,angle=0]{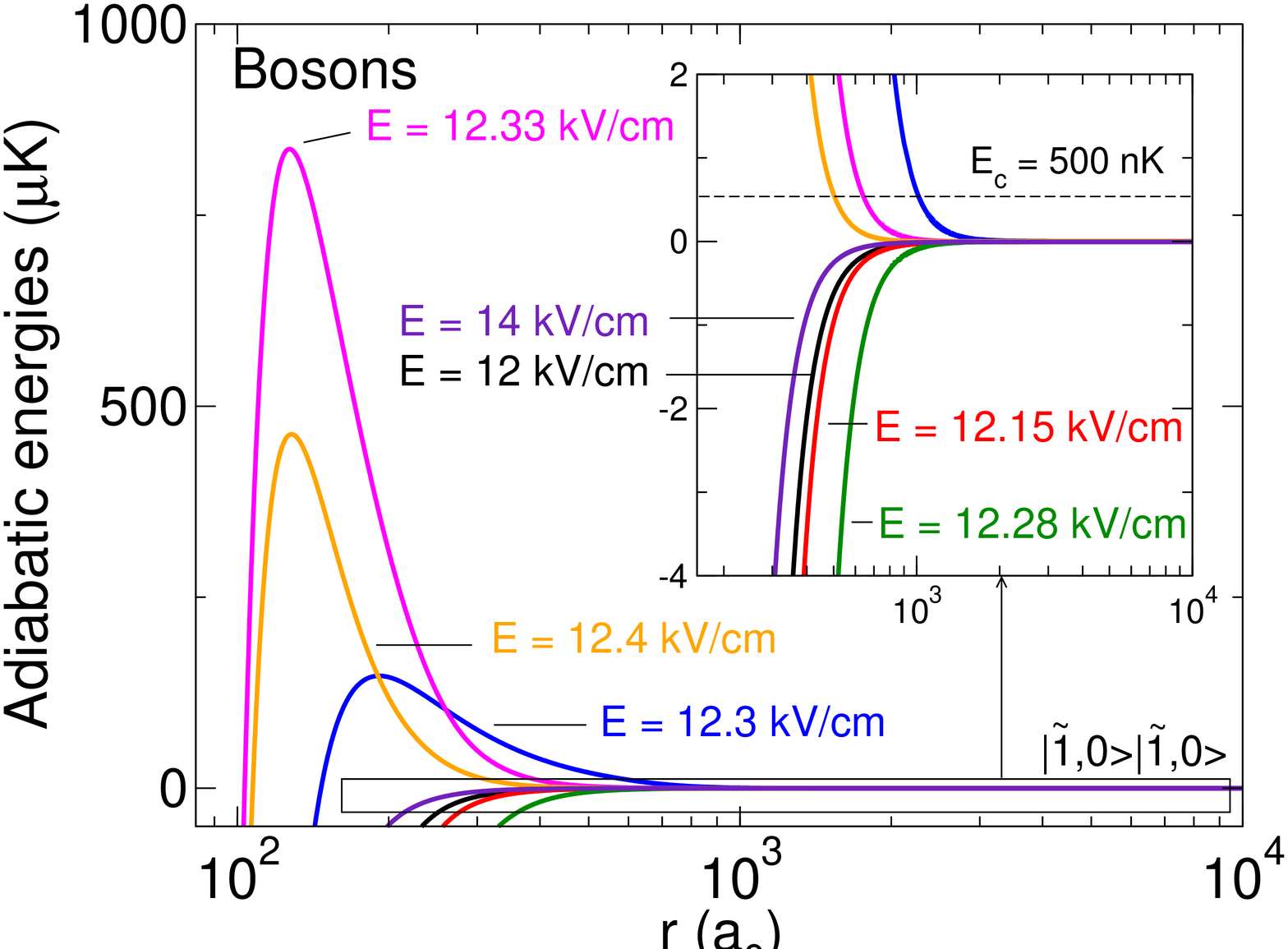} \\
\includegraphics*[width=9cm,keepaspectratio=true,angle=0]{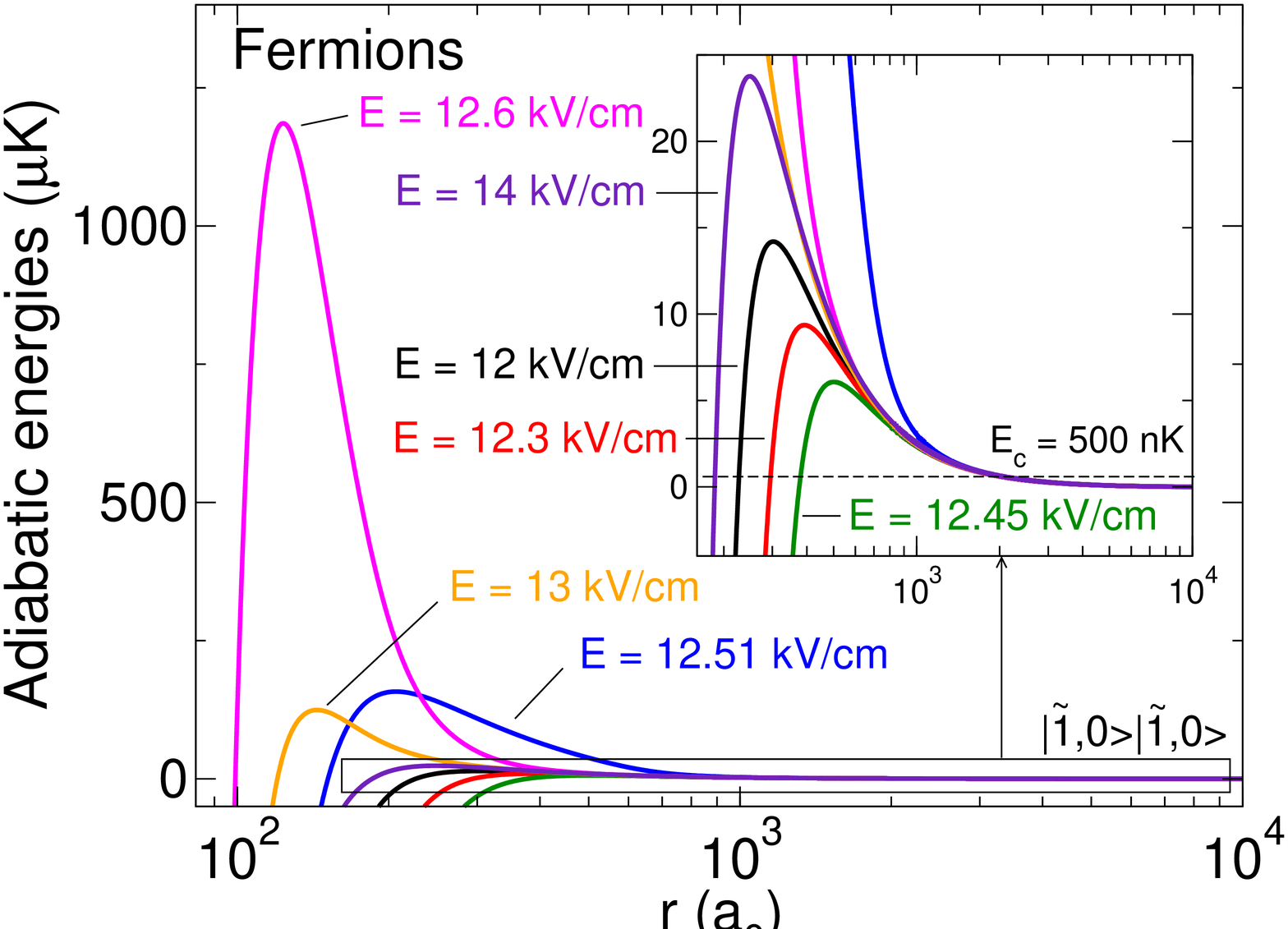}
\end{center}
\caption{(Color online) Lowest incident channel taken from the adiabatic energies for the initial colliding state $|\tilde{1},0\rangle$+$|\tilde{1},0\rangle$ at various electric fields (relative to the initial threshold energy). Upper panel: bosonic $^{41}$K$^{87}$Rb, s-wave dominated curve. Lower panel: fermionic $^{40}$K$^{87}$Rb, p-wave dominated curve. The green and blue curves represents the barriers just below and above the crossing of the threshold $|\tilde{1},0\rangle$+$|\tilde{1},0\rangle$ with $|\tilde{0},0\rangle$+$|\tilde{2},0\rangle$. The inset also indicates the collision energy $E_c=500$~nK.}
\label{BARRIER-FIG}
\end{figure}

The main interesting feature of Fig.~\ref{RATEN1-FIG}
comes from the presence of two sharply varying structures for the rates near $E_2^* \sim 11.5$~kV/cm and  $E_3^* \sim 12.5$~kV/cm (the later is shown in the inset (b)) and two smoother ones near $E_1^* \sim 10.5$~kV/cm and $E_4^* \sim 27$~kV/cm (the former is shown in the inset (c)).
The blue boxes of Fig.~\ref{NRG-2PLE-FIG} indicates the position of
the electric fields $E^*_1,E^*_2,E^*_3$ 
where other coupled combined states,
respectively $|\tilde{0},0\rangle+|\tilde{2},\pm2\rangle$,
$|\tilde{0},0\rangle+|\tilde{2},\pm1\rangle$ and
$|\tilde{0},0\rangle+|\tilde{2},0\rangle$,
cross the threshold of our initial combined state
$|\tilde{1},0\rangle+|\tilde{1},0\rangle$. 
The fourth structure occurs when the combined states
$|\tilde{1},\pm1\rangle+|\tilde{2},\pm2\rangle$ or 
$|\tilde{1},\pm1\rangle+|\tilde{2},\mp2\rangle$
cross the threshold of $|\tilde{1},0\rangle+|\tilde{1},0\rangle$ (green box in Fig.~\ref{NRG-2PLE-FIG}). 
Note that in our case the coupled combined state always starts
above the initial state and finishes below it as $E$ increases.
The structure near $E^*_1$ is far less pronounced than the two others near $E^*_2$ and $E^*_3$. The reason is
that there is no direct dipole-dipole coupling between the combined molecular states 
$|\tilde{0},0\rangle+|\tilde{2},\pm2\rangle$ and $|\tilde{1},0\rangle+|\tilde{1},0\rangle$
since a change in $\Delta m_{n_2} = m_{n_2}'- m_{n_2} = \pm 2$ is not allowed 
from the 3-j symbol of Eq.~\ref{VMULT}. 
Those two combined states are only coupled to second order, weaker than a direct coupling.
The same explanation holds for the structure near $E^*_4$.
For the two other combined states 
$|\tilde{0},0\rangle+|\tilde{2},\pm1\rangle$ and 
$|\tilde{0},0\rangle+|\tilde{2},0\rangle$ near $E^*_2$ and $E^*_3$, 
a direct dipole-dipole coupling occurs with $|\tilde{1},0\rangle+|\tilde{1},0\rangle$
since a change in $\Delta m_{n_2} = 0, \pm 1$ is allowed.
We want to emphasize that these structures are by no means resonances mediated by some certain tetramer bound-states. The condition of full loss probability at short range insures that all possible resonances are washed out in the numerical calculation. In other words here, the rate coefficients correspond only to some background scattering processes and not to some resonant scattering ones.

The principle of this mechanism was originally explained by Avdeenkov et al.~\cite{Avdeenkov_PRA_73_022707_2006} for non-lossy molecules. 
The basic idea is the following. We consider the initial colliding state of interest, here $|\tilde{1},0\rangle+|\tilde{1},0\rangle$ and we take the structure near $E^*_3$ in the rates (inset (b) of Fig.~\ref{RATEN1-FIG}) as an example. When the coupled state
$|\tilde{0},0\rangle+|\tilde{2},0\rangle$ is above the initial state,
the corresponding van der Waals interaction for the initial state will be attractive.
This can be seen from a second order perturbation theory as explained in Ref.~\cite{Stuhl_N_492_396_2012}.
Contributions from other states above or below can occur, however it is a good approximation to consider that the two combined states are isolated, when the two levels are about to cross.
The adiabatic curves are displayed in Fig.~\ref{SPAG-FIG} for the two combined states $|\tilde{1},0\rangle+|\tilde{1},0\rangle$ and
$|\tilde{0},0\rangle+|\tilde{2},0\rangle$, at $E < E^*_3$.
The lowest incident channel is displayed in bold.
When the electric field is tuned and the energy of $|\tilde{0},0\rangle+|\tilde{2},0\rangle$ approaches the one of $|\tilde{1},0\rangle+|\tilde{1},0\rangle$, the strength of the long-range attractive van der Waals interaction is increased. When the coupled  state crosses the initial state from above to below, the sign of the van der Waals interaction changes from attractive to repulsive.
This is shown in Fig.~\ref{BARRIER-FIG} for bosons (upper panel) and fermions (lower panel) where the  lowest incident channels are modified by the electric field tuning. Far from $E^*_3$, at $E=12$~kV/cm (black), we have the usual s-wave dominating behaviour for the bosons and the  p-wave dominating one for the fermions (see insets). When $E \to E^*_3$ slightly below $E^*_3$, for example at $E=12.28$~kV/cm for bosons and $E=12.45$~kV/cm for fermions (green), the incident curves get more attractive as the strength of the van der Waals interaction increases. Slightly above $E^*_3$ at $E=12.3$~kV/cm for bosons and $E=12.51$~kV/cm for fermions (blue), no matter the bosonic or fermionic symmetry character, the incident channels become repulsive due to the change of sign of the van der Waals interaction. At $E=12.33$~kV/cm for bosons and $E=12.6$~kV/cm for fermions, the barrier is quite high around 1~mK compared to the collision energy
of 500~nK. When the electric field further increases (orange, indigo) far from $E^*_3$ the barriers decrease and go back to a similar situation when $E < E^*_3$.
This explains why the rates sharply increase when $E$ is slightly below $E^*_3$ and sharply decrease when $E$ is slightly above $E^*_3$. 
Also it confirms again that this is a background scattering process where the value of the rates are mediated by the change of the barrier of the incident channels only. The sharp, quick-varying structures come
from the small range of electric field where the crossings of the combined states take place.
This mechanism is more striking for bosons because we can see the rise of the repulsive barrier for the lowest s-wave dominated incident channel. \\

For this particular system KRb at this particular collision energy, we show the contribution of the loss process compared to the elastic one in the insets (b) of Fig.~\ref{RATEN1-FIG}. We see that, around the crossing of the two combined states, one can invert the dominant loss process in favour to the elastic one. However, a small ratio of elastic over loss rates $\gamma$ is found, $\gamma=7$ for the bosons and $\gamma=20$ for the fermions, far from the ideal values of $\sim$1000 to perform efficient evaporative cooling.
Near and at threshold $E = E^*_3$ of the crossing states, the inelastic process when it is open by the collision energy (see insets (b) of Fig.~\ref{RATEN1-FIG}) is enhanced for the state-to-state transition $|\tilde{1},0\rangle+|\tilde{1},0\rangle$ $\to$ $|\tilde{0},0\rangle+|\tilde{2},0\rangle$ because of the strong long-range mixing between the two coupled combined states. This takes place in a small range of electric field $E=[12.3-12.4]$~kV/cm for bosons and $E=[12.5-12.6]$~kV/cm for fermions. This is the price to pay to get a repulsive barrier and a decrease of the reactive rates at short range above the threshold $E > E^*_3$.
On one hand reactive processes are protected at short range but on the other hand inelastic processes are increased due to the coupling with the state that is responsible for the repulsive incident barrier. It is then a compromise between those two processes for the loss processes which probably depends on the properties of the molecular system such as the dipole moment, rotational constant, electronic van der Waals coefficient and mass.

\section{Conclusion}

We studied the ultracold collisions of rotationally excited polar molecules in the presence of an electric field, taking KRb as an example, including the electric dipole, quadrupole and octopole moments in the expression of the long-range multipole-multipole interaction.
The consideration of quadrupoles and octopoles does not affect the incident channel mediated processes like elastic and reactive/sticky collisions at ultralow energies, however it affects the inelastic processes which are mediated by state-to-state transitions.
Elastic, inelastic and reactive/sticky rates can be tuned when two combined molecular states are brought close together with the electric field, and even invert the dominant loss process in favour to the elastic one. This tunability of the rates is created without the presence of any bound-state resonances which implies that it can work for highly lossy (including reactive or sticky) collisional processes. This is found either for bosons and fermions and for non-confined, free space collisions.
Even though the mechanism is quite tunable, we cannot create large value of the ratio of elastic over loss process to perform efficient evaporative cooling for the present KRb + KRb system. Therefore, future studies will involve other polar molecules with higher electric dipole moments to prospect if larger ratio can be reached with this mechanism.

\section*{Acknowledgments}

This work was performed under the financial support of the COPOMOL project
\# ANR-13-IS04-0004 from Agence Nationale de la Recherche. We also acknowledge financial support from Projet Attractivit\'e 2014 from Universit\'e Paris-Sud. We thank Maxence Lepers, Maykel L. Gonz{\'a}lez-Mart{\'i}nez, Olivier Dulieu and John L. Bohn for discussions.

\section*{Appendix A: Collisions of $\rm{K}\rm{Rb}(|\tilde{0},0\rangle) +  \rm{K}\rm{Rb}(|\tilde{0},0\rangle)$}

\begin{figure} [h]
\begin{center}
\includegraphics*[width=6cm,keepaspectratio=true,angle=-90]{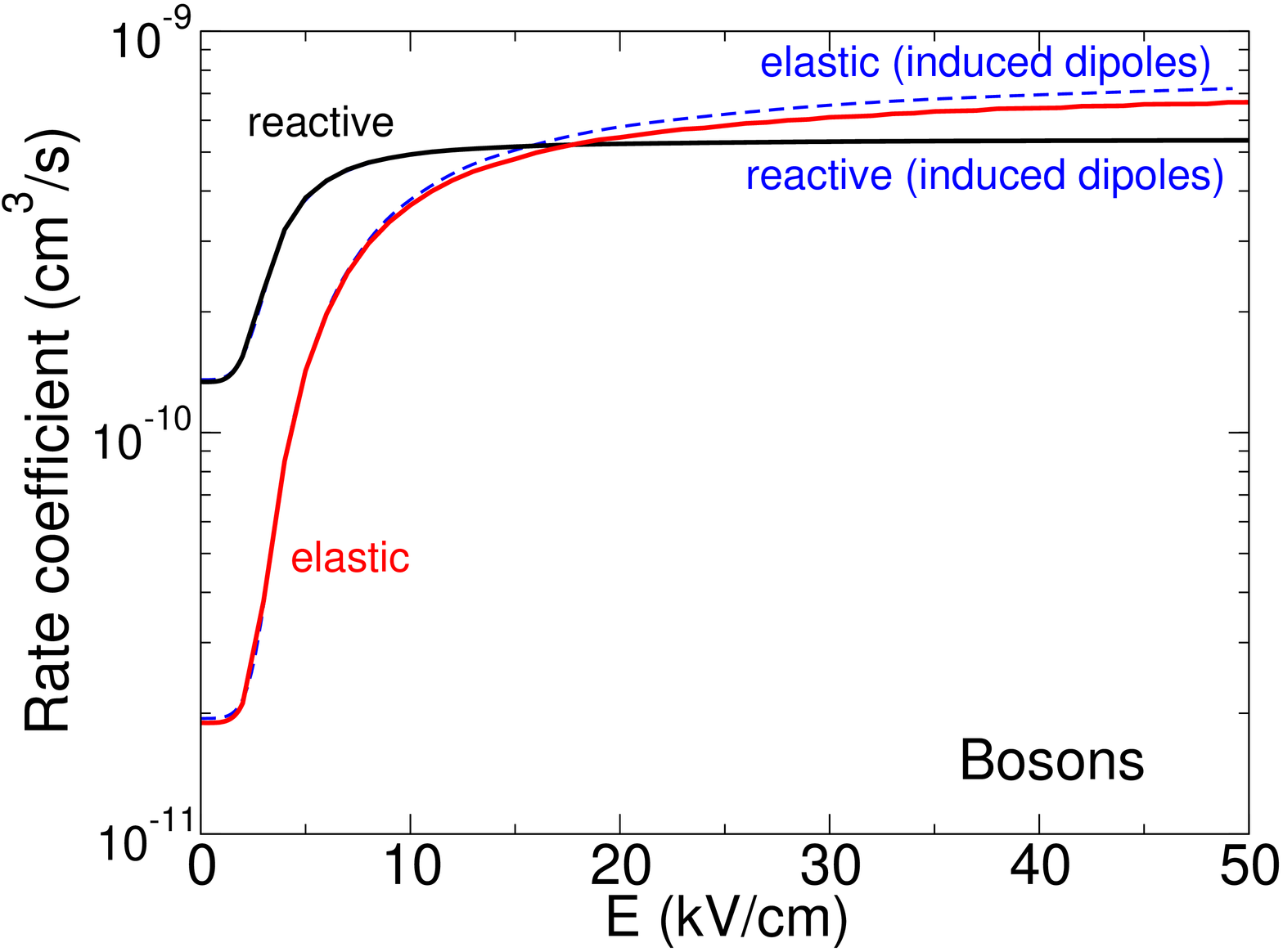}
\includegraphics*[width=6cm,keepaspectratio=true,angle=-90]{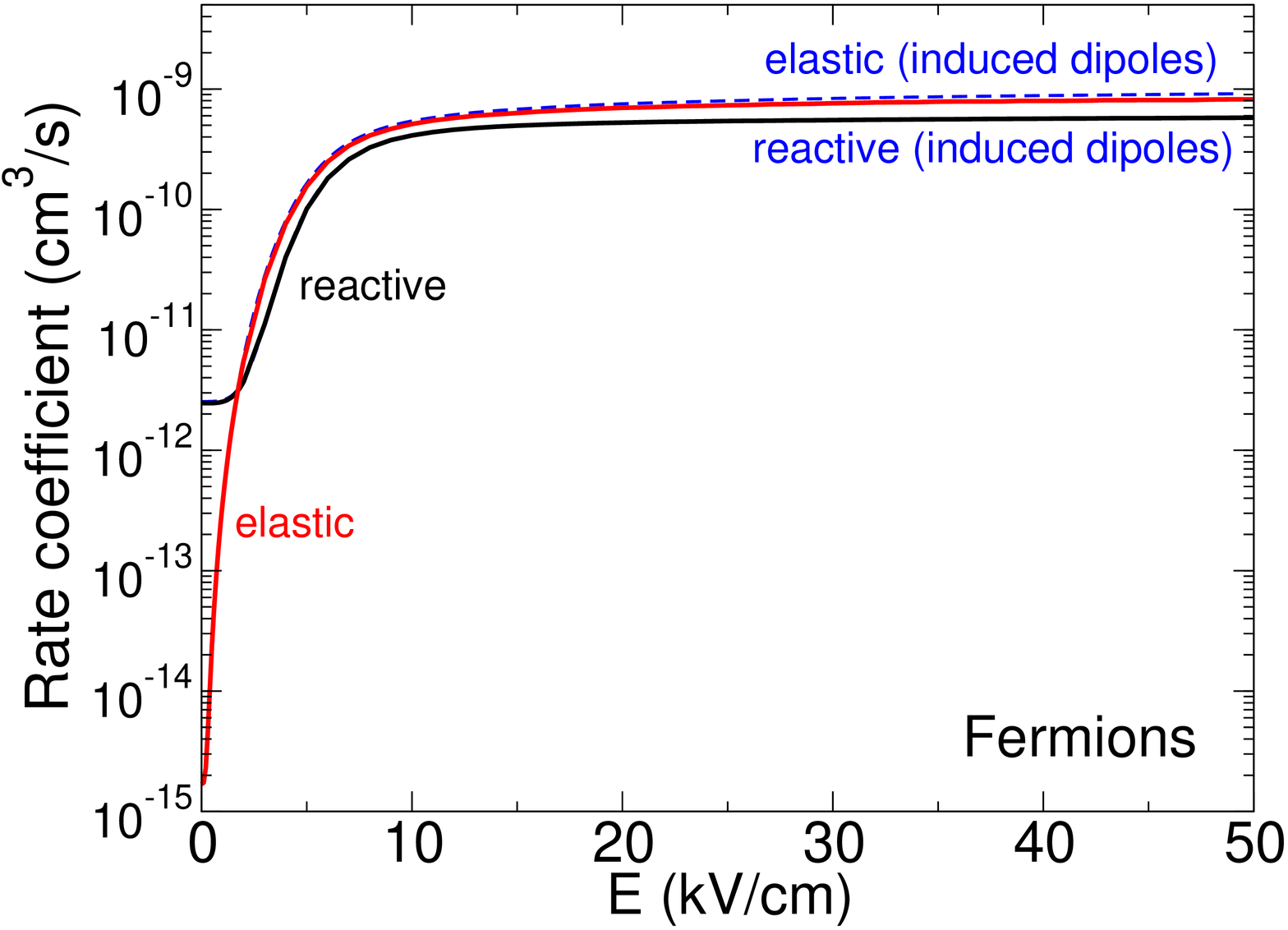}
\end{center}
\caption{(Color online) Rate coefficient for the initial colliding state $|\tilde{0},0\rangle$+$|\tilde{0},0\rangle$ at various electric fields. The collision energy is fixed to $E_c=500$~nK. Left panel: bosonic $^{41}$K$^{87}$Rb. Right panel: fermionic $^{40}$K$^{87}$Rb. Reactive: black curve. Elastic: red curve. The calculations using the rotational structure of the molecules are compared with a simpler calculation using only the induced dipole moments of the molecules (blue dashed curves). The reactive curves cannot be graphically distinguished.}
\label{RATEN0-FIG}
\end{figure}

As a check of our numerical formalism including the internal rotational states, we consider collisions of KRb molecules in the ground states $\tilde{n}=0, m_n=0$. This has been extensively studied in the past
\cite{Quemener_PRA_81_022702_2010,Quemener_PRA_83_012705_2011,Julienne_PCCP_13_19114_2011,Quemener_PRA_84_062703_2011}.
It was found that reactive rate coefficients behave as $d_{\rm{ind}}^2$ for
bosons~\cite{Quemener_PRA_84_062703_2011} and $d_{\rm{ind}}^6$~\cite{Quemener_PRA_81_022702_2010} for fermions while for both, elastic rate coefficients behave as $d_{\rm{ind}}^4$ ~\cite{Bohn_NJP_11_055039_2009}.
These former studies never included explicitly the rotational structure of the molecules
because for these ground state molecules,
one can circumvent the full rotational problem for loss collisions with a simpler one
(see for example the appendix of Ref.~\cite{Quemener_PRA_83_012705_2011})
using two effective induced dipole moments.
We present in Fig.~\ref{RATEN0-FIG} the rate coefficients for bosons  and fermions
for two molecules in the ground rotational state, including the rotational structure of the molecules.
The dashed blue lines represent the rates computed with induced dipole
moments~\cite{Quemener_PRA_81_022702_2010,Quemener_PRA_84_062703_2011}.
For the reactive rates the curves are graphically identical with the full rotational calculations and overlap each others. We confirm here that the induced dipole moment method is a very good approximation for reactive rates for the ground rotational state.
For the elastic rates, while at low electric fields the curves are graphically identical, there is a slight difference at higher electric fields ($E\sim 10$~kV/cm and more).
A qualitative explanation could come from the couplings with locally-open channels at short range corresponding to closed channels of higher rotational states at long range. This slightly modifies the propagated log-derivative of the incident channel. These couplings are taken into account in the full rotational method while for the induced dipole moment method they are not. As the electric field increases, these couplings with higher rotational states become stronger.
Those channels, when locally open, modify slightly the phase-shift of the incident channel.
But since they are asymptotically closed they do not contribute to the value of the elastic probability. As the reactive rates are proportional to the loss probability (equal to unity minus the elastic probability) they are not affected by the locally-open channels, while the elastic rates are proportional to the incident phase shift and are slightly modified by the locally-open channels.

\newpage

\section*{Appendix B: Rate coefficients convergence of bosonic $^{41}$K$^{87}$Rb +  $^{41}$K$^{87}$Rb collisions with inclusion of dipole (D), quadrupole (Q), octopole (O)}

\subsection*{$^{41}${\rm K}$^{87}${\rm Rb}$(|\tilde{0},0\rangle)$ {\rm +}  $^{41}${\rm K}$^{87}${\rm Rb}$(|\tilde{0},0\rangle)$, {\rm reactive rate (cm}$^3${\rm /s)}}

\begin{table}[h]
\centering
\begin{tabular}{c|c|c|c}
\hline
  electric field (kV/cm)    & D &   DQ    &   DQO    \\
\hline
    0  &    $1.34\times10^{-10}$  &   $1.28\times10^{-10}$    &   $1.32\times10^{-10}$       \\
    \hline
    10  &   $4.93\times10^{-10}$  &   $4.82\times10^{-10}$    &   $5.10\times10^{-10}$       \\
\hline
    20  &   $5.24\times10^{-10}$  &   $5.30\times10^{-10}$   &   $5.26\times10^{-10}$       \\
\hline
    30  &   $5.31\times10^{-10}$  &   $5.33\times10^{-10}$     &   $5.28\times10^{-10}$   \\
\hline
    40  &   $5.34\times10^{-10}$  &   $5.37\times10^{-10}$     &   $5.28\times10^{-10}$    \\
\hline
    50  &   $5.35\times10^{-10}$  &   $5.35\times10^{-10}$     &   $5.33\times10^{-10}$    \\
\hline
\end{tabular}
\end{table}

\subsection*{$^{41}${\rm K}$^{87}${\rm Rb}$(|\tilde{0},0\rangle)$ {\rm +}  $^{41}${\rm K}$^{87}${\rm Rb}$(|\tilde{0},0\rangle)$, {\rm elastic rate (cm}$^3${\rm /s)}}

\begin{table}[h]
\centering
\begin{tabular}{c|c|c|c}
\hline
  electric field (kV/cm)    & D &   DQ    &   DQO    \\
\hline
    0  &   $1.89\times10^{-11}$  &   $1.81\times10^{-11}$    &     $1.77\times10^{-11}$       \\
    \hline
    10  &   $3.69\times10^{-10}$  &   $3.61\times10^{-10}$     &   $4.00\times10^{-10}$       \\
\hline
    20  &   $5.44\times10^{-10}$  &   $5.69\times10^{-10}$       & $5.41\times10^{-10}$       \\
\hline
    30  &   $6.11\times10^{-10}$  &   $6.11\times10^{-10}$     &   $6.40\times10^{-10}$    \\
\hline
    40  &   $6.43\times10^{-10}$  &   $6.73\times10^{-10}$     &   $6.40\times10^{-10}$     \\
\hline
    50  &   $6.66\times10^{-10}$  &   $6.54\times10^{-10}$     &   $6.50\times10^{-10}$   \\
\hline
\end{tabular}
\end{table}

\subsection*{$^{41}${\rm K}$^{87}${\rm Rb}$(|\tilde{1},0\rangle)$ {\rm +}  $^{41}${\rm K}$^{87}${\rm Rb}$(|\tilde{1},0\rangle)$, {\rm reactive rate (cm}$^3${\rm /s)}}

\begin{table}[h]
\centering
\begin{tabular}{c|c|c|c}
\hline
  electric field (kV/cm)    & D &   DQ    &   DQO    \\
\hline
    0  &   $1.32\times10^{-10}$  &   $1.31\times10^{-10}$    &   $1.28\times10^{-10}$       \\
    \hline
    10  &   $1.68\times10^{-10}$  &   $1.69\times10^{-10}$    &   $1.67\times10^{-10}$     \\
\hline
    20  &   $1.30\times10^{-10}$  &   $1.31\times10^{-10}$     &   $1.32\times10^{-10}$      \\
\hline
    30  &   $1.93\times10^{-10}$  &   $1.91\times10^{-10}$     &   $1.93\times10^{-10}$    \\
\hline
    40  &   $3.45\times10^{-10}$  &   $3.45\times10^{-10}$      &   $3.46\times10^{-10}$     \\
\hline
    50  &   $4.15\times10^{-10}$  &   $4.21\times10^{-10}$     &   $4.10\times10^{-10}$    \\
\hline
\end{tabular}
\end{table}

\newpage

\subsection*{$^{41}${\rm K}$^{87}${\rm Rb}$(|\tilde{1},0\rangle)$ {\rm +}  $^{41}${\rm K}$^{87}${\rm Rb}$(|\tilde{1},0\rangle)$, {\rm elastic rate (cm}$^3${\rm /s)}}

\begin{table}[h]
\centering
\begin{tabular}{c|c|c|c}
\hline
  electric field (kV/cm)    & D &   DQ    &   DQO    \\
\hline
    0   &   $1.85\times10^{-11}$  &   $1.94\times10^{-11}$    &   $1.97\times10^{-11}$       \\
\hline
    10  &   $2.43\times10^{-11}$  &   $2.44\times10^{-11}$   &   $2.43\times10^{-11}$       \\
\hline
    20  &   $1.78\times10^{-11}$  &   $1.81\times10^{-11}$     &   $1.83\times10^{-11}$       \\
\hline
    30  &   $2.81\times10^{-11}$  &   $2.79\times10^{-11}$      &   $2.84\times10^{-11}$    \\
\hline
    40  &   $1.04\times10^{-10}$  &   $1.04\times10^{-10}$     &   $1.04\times10^{-10}$    \\
\hline
    50  &   $1.83\times10^{-10}$  &   $1.86\times10^{-10}$    &   $1.76\times10^{-10}$   \\
\hline
\end{tabular}
\end{table}

\subsection*{$^{41}${\rm K}$^{87}${\rm Rb}$(|\tilde{1},0\rangle)$ {\rm +}  $^{41}${\rm K}$^{87}${\rm Rb}$(|\tilde{1},0\rangle)$, {\rm inelastic rate (cm}$^3${\rm /s)}}

\begin{table}[h]
\centering
\begin{tabular}{c|c|c|c}
\hline
  electric field (kV/cm)    & D &   DQ    &   DQO    \\
\hline
    0  &   $9.73\times10^{-15}$  &   $1.38\times10^{-14}$    &   $6.18\times10^{-14}$       \\
    \hline
    10  &   $1.66\times10^{-16}$  &   $1.54\times10^{-13}$  &   $1.62\times10^{-13}$       \\
\hline
    20  &   $1.37\times10^{-16}$  &   $8.51\times10^{-14}$   &   $9.69\times10^{-14}$       \\
\hline
    30  &   $1.76\times10^{-16}$  &   $1.44\times10^{-13}$    &   $1.94\times10^{-13}$     \\
\hline
    40  &   $3.28\times10^{-16}$  &   $1.04\times10^{-13}$   &   $9.82\times10^{-14}$      \\
\hline
    50  &   $4.28\times10^{-16}$  &   $2.38\times10^{-13}$   &   $1.86\times10^{-13}$     \\
\hline
\end{tabular}
\end{table}

\section*{Appendix C: Rate coefficients convergence of fermionic $^{40}$K$^{87}$Rb +  $^{40}$K$^{87}$Rb collisions with inclusion of dipole (D), quadrupole (Q), octopole (O)}

\subsection*{$^{40}${\rm K}$^{87}${\rm Rb}$(|\tilde{0},0\rangle)$ {\rm +}  $^{40}${\rm K}$^{87}${\rm Rb}$(|\tilde{0},0\rangle)$, {\rm reactive rate (cm}$^3${\rm /s)}}

\begin{table}[h]
\centering
\begin{tabular}{c|c|c|c}
\hline
  electric field (kV/cm)    & D &   DQ    &   DQO    \\
\hline
    0  &   $2.46\times10^{-12}$  &   $2.52\times10^{-12}$    &   $2.57\times10^{-12}$       \\
    \hline
    10  &   $4.12\times10^{-10}$  &   $4.24\times10^{-10}$    &   $3.88\times10^{-10}$      \\
\hline
    20  &   $5.26\times10^{-10}$  &   $5.18\times10^{-10}$      &   $5.22\times10^{-10}$       \\
\hline
    30  &   $5.52\times10^{-10}$  &   $5.52\times10^{-10}$     &   $5.43\times10^{-10}$    \\
\hline
    40  &   $5.67\times10^{-10}$  &   $5.62\times10^{-10}$      &   $5.65\times10^{-10}$     \\
\hline
    50  &   $5.78\times10^{-10}$  &   $5.77\times10^{-10}$     &   $5.76\times10^{-10}$    \\
\hline
\end{tabular}
\end{table}

\newpage

\subsection*{$^{40}${\rm K}$^{87}${\rm Rb}$(|\tilde{0},0\rangle)$ {\rm +}  $^{40}${\rm K}$^{87}${\rm Rb}$(|\tilde{0},0\rangle)$, {\rm elastic rate (cm}$^3${\rm /s)}}

\begin{table}[h]
\centering
\begin{tabular}{c|c|c|c}
\hline
  electric field (kV/cm)    & D &   DQ    &   DQO    \\
\hline
    0  &   $1.68\times10^{-15}$  &   $1.81\times10^{-15}$    &   $1.76\times10^{-15}$       \\
    \hline
    10  &   $5.12\times10^{-10}$  &   $5.37\times10^{-10}$    &   $5.00\times10^{-10}$      \\
\hline
    20  &   $6.98\times10^{-10}$  &   $7.05\times10^{-10}$     &   $6.82\times10^{-10}$      \\
\hline
    30  &   $7.62\times10^{-10}$  &   $7.60\times10^{-10}$      &   $8.07\times10^{-10}$   \\
\hline
    40  &   $7.98\times10^{-10}$  &   $8.17\times10^{-10}$    &   $8.25\times10^{-10}$     \\
\hline
    50  &   $8.21\times10^{-10}$  &   $8.06\times10^{-10}$    &   $7.98\times10^{-10}$   \\
\hline
\end{tabular}
\end{table}

\subsection*{$^{40}${\rm K}$^{87}${\rm Rb}$(|\tilde{1},0\rangle)$ {\rm +}  $^{40}${\rm K}$^{87}${\rm Rb}$(|\tilde{1},0\rangle)$, {\rm reactive rate (cm}$^3${\rm /s)}}

\begin{table}[h]
\centering
\begin{tabular}{c|c|c|c}
\hline
  electric field (kV/cm)    & D &   DQ    &   DQO    \\
\hline
    0  &   $2.38\times10^{-12}$  &   $2.81\times10^{-12}$    &   $2.83\times10^{-12}$       \\
    \hline
    10  &   $5.14\times10^{-12}$  &   $5.06\times10^{-12}$     &   $5.11\times10^{-12}$        \\
\hline
    20  &   $2.21\times10^{-12}$  &   $2.22\times10^{-12}$      &   $2.21\times10^{-12}$       \\
\hline
    30  &   $6.26\times10^{-12}$  &   $6.26\times10^{-12}$    &   $6.23\times10^{-12}$    \\
\hline
    40  &   $5.48\times10^{-11}$  &   $5.51\times10^{-11}$     &   $5.54\times10^{-11}$      \\
\hline
    50  &   $1.57\times10^{-10}$  &   $1.55\times10^{-10}$     &   $1.54\times10^{-10}$    \\
\hline
\end{tabular}
\end{table}

\subsection*{$^{40}${\rm K}$^{87}${\rm Rb}$(|\tilde{1},0\rangle)$ {\rm +}  $^{40}${\rm K}$^{87}${\rm Rb}$(|\tilde{1},0\rangle)$, {\rm elastic rate (cm}$^3${\rm /s)}}

\begin{table}[h]
\centering
\begin{tabular}{c|c|c|c}
\hline
  electric field (kV/cm)    & D &   DQ    &   DQO    \\
\hline
    0  &   $1.53\times10^{-15}$  &   $ 1.71\times10^{-15}$    &   $1.78\times10^{-15}$       \\
    \hline
    10  &   $9.00\times10^{-12}$  &   $8.99\times10^{-12}$    &   $9.00\times10^{-12}$       \\
\hline
    20  &   $3.94\times10^{-15}$  &   $3.99\times10^{-15}$      &  $3.97\times10^{-15}$       \\
\hline
    30  &   $1.44\times10^{-11}$  &   $1.44\times10^{-11}$     &   $1.44\times10^{-11}$    \\
\hline
    40  &   $9.59\times10^{-11}$  &   $9.58\times10^{-11}$     &   $9.57\times10^{-11}$     \\
\hline
    50  &   $2.21\times10^{-10}$  &   $2.25\times10^{-10}$     &   $2.22\times10^{-10}$   \\
\hline
\end{tabular}
\end{table}

\newpage

\subsection*{$^{40}${\rm K}$^{87}${\rm Rb}$(|\tilde{1},0\rangle)$ {\rm +}  $^{40}${\rm K}$^{87}${\rm Rb}$(|\tilde{1},0\rangle)$, {\rm inelastic rate (cm}$^3${\rm /s)}}

\begin{table}[h]
\centering
\begin{tabular}{c|c|c|c}
\hline
  electric field (kV/cm)    & D &   DQ    &   DQO    \\
\hline
    0  &   $1.61\times10^{-17}$  &   $9.84 \times10^{-17}$    &   $4.38\times10^{-16}$       \\
    \hline
    10  &   $5.08\times10^{-18}$  &   $4.28\times10^{-15}$  &   $4.78\times10^{-15}$        \\
\hline
    20  &   $2.10\times10^{-18}$  &   $1.35\times10^{-15}$   &   $1.37\times10^{-15}$       \\
\hline
    30  &   $4.50\times10^{-18}$  &   $4.54\times10^{-15}$   &   $6.51\times10^{-15}$     \\
\hline
    40  &   $4.05\times10^{-17}$  &   $2.27\times10^{-14}$  &   $1.69\times10^{-14}$      \\
\hline
    50  &   $1.35\times10^{-16}$  &   $6.59\times10^{-14}$  &   $8.35\times10^{-14}$     \\
\hline
\end{tabular}
\end{table}

\end{document}